\documentclass[twocolumn,showpacs,preprintnumbers,amsmath,amssymb]{revtex4}

\usepackage{CJK}
\usepackage{graphicx}
\usepackage{dcolumn}
\usepackage{bm}

\def\btbl{\begin{tabular}} \def\etbl{\end{tabular}}
\def\bcc{\begin{center}} \def\ecc{\end{center}}
\def\beq{\begin{equation}} \def\eeq{\end{equation}}
\def\btbl{\begin{tabular}} \def\etbl{\end{tabular}}

\def\E941{{\footnotesize E941}} 
\def\NA49{{\footnotesize NA49}} \def\NA35{{\footnotesize NA35}}

\begin{document}
\title{Cold nuclear matter effects on the color singlet $J/\psi$ production \\
in $dAu$ collisions at RHIC}

\author{Ze-Fang  Jiang$^{1}$}
\author{Sheng-Qin Feng$^{1,2,4}$}
\email{fengsq@ctgu.edu.cn}
\author{Zhong-Bao Yin$^{2,3}$}
\author{Ya-Fei Shi$^{1}$}
\author{Xian-Bao Yuan$^{1}$}

\affiliation{$^1$ College of Science, China Three Gorges University£¬Yichang 443002, China}
\affiliation{$^2$ Key Laboratory of Quark and Lepton Physics (Huazhong Normal Univ.), Ministry of Education£¬Wuhan 430079£¬China}
\affiliation{$^3$  Institute of Particle Physics, Central China Normal University, Wuhan 430079, China}
\affiliation{$^4$ School of Physics and Technology, Wuhan University, Wuhan 430072, China }

\begin{abstract}
We use a Modified DKLMT model (called M-DKLMT model) to study the cold nuclear matter (CNM) effects on the color singlet $J/\psi$ production in $dAu$
collisions at RHIC. The cold nuclear effect of dipole-nucleus interactions has been investigated by introducing a nuclear geometric
effect function $f(\xi)$ to study the nuclear geometry distribution effect in relativistic heavy-ion collisions. The dependencies of nuclear modification
factors ($R_{dA}$) on rapidity and centrality are studied and compared to experimental data. It is found that the M-DKLMT model can well describe the
experimental results at both forward- and mid-rapidity regions in $dAu$ collisions at RHIC.\\

\end{abstract}


\maketitle

\section{Introduction}
\label{intro}
Heavy quark production in high-energy nuclear collisions has been a focus of interest for many years. Heavy quarks are essential probes of the
evolution of the medium created in heavy-ion collisions since they are produced predominantly in the early stage of nuclear collisions \cite{lab1}.
Heavy-quark production in $pp$ collisions was studied not only to test the  perturbative
quantum chromo-dynamics but also to serve as a baseline for studying heavy-ion collisions \cite{lab2,lab3,lab4}.
Although suppression of high $p_{T}$  particles was predicted as an effect of parton energy loss in the hot dense medium created
in relativistic heavy-ion collisions \cite{lab5,lab6,lab7},  it is difficult to account for the comparable suppression of heavy flavors to
that of light flavors solely with hot nuclear matter effects \cite{lab8}. To understand comprehensively the parton energy loss mechanism in
hot dense medium,  it is essential to explore fully the underlying cold nuclear matter (CNM) effects.

While measurement of heavy-flavor production in elementary collisions is crucial for test of the validity of the current theoretical framework and for
inputs to phenomenological models to describe heavy-flavor production in nucleus-nucleus collisions, control experiments with $pA$ or $dA$ collisions allow us
to probe those CNM effects. These include modifications of the parton distribution function (PDF) and $k_{T}$  broadening, with minimal impact from the
hot nuclear medium. Because heavy quarks are produced primarily by gluon fusion at RHIC, modification of the gluon density in nucleus can be observed
in charm and bottom production rates \cite{lab9,lab10}.

Basing on the McLerran-Venugopalan (MV) model \cite{lab11}, Dominguez, Kharzeev, Levin, Mueller and Tuchin (DKLMT) proposed a model \cite{lab12} to analyze the
gluon saturation effects on the color singlet $J/\psi$ productions in $dA$  and $AA$  collisions at RHIC energies \cite{lab13,lab14,lab15,lab16}.
The DKLMT model \cite{lab12} assumes that $c\bar{c}$ pair in a color-octet state propagates through the nucleus and becomes color singlet inside the nucleus.
In the large $N_{c}$ approximation further color conversions of the  $c\bar{c}$ state are suppressed and thus can be neglected. Therefore,
in this case the $c\bar{c}$ experiences the last inelastic interaction
inside the nucleus after which it re-scatters only elastically. Additionally, DKLMT model treats the $J/\psi$ wave function accurately with parameters
determined from a fit to the exclusive $J/\psi$ production in deep inelastic scattering.

Basing on the DKLMT model, we propose a new form of cold nuclear matter effect on the color singlet $J/\psi$ production mechanism in order to describe $dA$
experimental data at RHIC. This Modified DKLMT model (M-DKLMT model) can describe the centrality and rapidity dependencies of nuclear modification
factor ($R_{dA}$) for $J/\psi$ productions in $dAu$ collisions at RHIC.

This paper is organized as follows: Section 2 introduces the M-DKLMT model including the nuclear modification effects in  $dAu$ collisions;
Section 3 is dedicated to the description of the numerical calculation performed with DHJ model \cite{lab17} for the dipole scattering amplitude, and
the calculation results are then compared with the experimental data at RHIC. The conclusions are summarized in Section 4.

\begin{figure*}[ht]
\centering \resizebox{0.80\textwidth}{!}{
\includegraphics{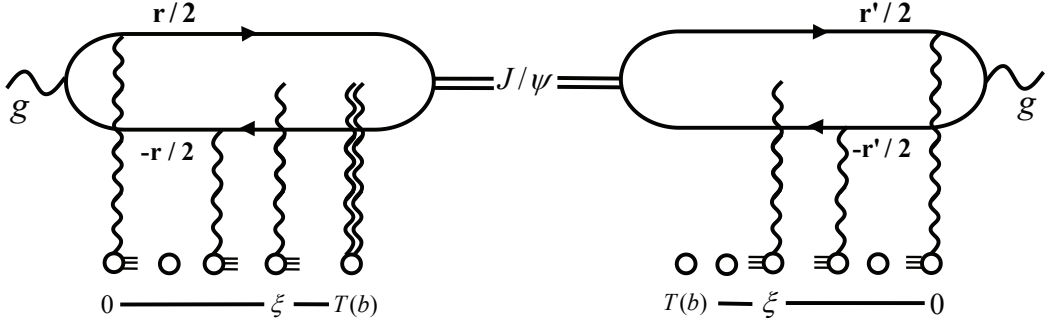}}
\caption{A sketch diagram for $gA \rightarrow J/\psi$ . The longitudinal coordinate $\xi$ is
the point where the last inelastic interaction takes place.}
\label{fig1}
\end{figure*}

\section{The M-DKLMT model and nuclear modification effect}
The DKLMT model \cite{lab12} contains three distinct assumptions.

(i)  In order to study $J/\psi$ production in high energy $pA$ (or $dA$) collisions, DKLMT model argued that $J/\psi$ production
in relativistic heavy-ion collisions should take into account the gluon saturation/color glass condensate effects. The $J/\psi$ production
cross section in high energy $pA$ (or $dA$) collisions can be written in the factorized form:

\begin{equation}
\frac{d\sigma_{pA\rightarrow J/\psi X}}{db^{2}dy}=x_{1}G(x_{1},m_{c}^{2})\frac{d\sigma_{gA\rightarrow J/\psi X}}{db^{2}}
\label{eq1}
\end{equation}

\noindent where a simple ansatz for the gluon distribution \cite{lab18} encoding the saturation \cite{lab19} is given as follows:
\begin{equation}  
       x_{1}G(x_{1},m_{c}^{2})=\left\{ \begin{array}{ll}
                              \frac{K}{\alpha_{s}(Q_{s})}m_{c}^{2}(1-x_{1})^{4}& \textrm{$, m_{c}<Q_{s}(x_{1})$}\\
                              \frac{K}{\alpha_{s}(Q_{s})}Q_{s}^{2}(1-x_{1})^{4}& \textrm{$, m_{c}>Q_{s}(x_{1})$}
                              \end{array}\right.
\label{eq:eq2} 
\end{equation}

\noindent with $x_{1}=(m_{c}/\sqrt{s})e^{y}$, $m_c$ is the charm quark mass and $\sqrt{s}$ the collision energy in the center of mass system, the normalization factor $K$
and $\alpha_{s}(Q_{s})$ are determined by a fit to $pp$ data and $d-Au$ data at RHIC.

(ii) In order to study $\frac{d\sigma_{gA\rightarrow J/\psi X}}{db^{2}}$ , DKLMT used a well developed phenomenology $\gamma A$ theory by stating a $\gamma p$ scattering
\begin{equation}
\frac{d\sigma_{\gamma A\rightarrow J/\psi X}}{dt}=\frac{1}{16} |A_{\gamma p\rightarrow J/\psi p}|^{2},
\label{eq3}
\end{equation}

\noindent with
\begin{eqnarray}
\label{eq4}
\begin{aligned}
& A_{\gamma p\rightarrow J/\psi X}(x,\Delta)=\\
&\int d^{2}be^{-i\Delta \cdot b}\int\limits_{0}^{1}dz\int \frac{d^{2}r}{4\pi}(\Psi_{J/\psi}^{\ast}\Psi_{\gamma})2i[1-S(x,\mathbf{r},\mathbf{b})]
\end{aligned}
\end{eqnarray}

\noindent where $t=-\Delta^{2}$ is the momentum transfer, and $\Psi_{J/\psi}^{\ast}\Psi_{\gamma}$=$\Phi_{\gamma}(\mathbf{r},z)$
with
\begin{equation}
\begin{aligned}
\phi_{\gamma}(\mathbf{r},z)&=\frac{2}{3}e\frac{N_{c}}{\pi}\{m_{c}^{2}K_{0}(m_{c}\mathbf{r})\phi_{T}(\mathbf{r},z) \\
&-[z^{2}+(1-z)^{2}]m_{c}K_{1}(m_{c}\mathbf{r})\partial_{r}\phi_{T}(\mathbf{r},z)\}
\label{eq5}
\end{aligned}
\end{equation}

\noindent where
\begin{equation}
\phi_{T}(r,z)=N_{T}z(1-z)exp[-\frac{r^{2}}{2R_{T}^{2}}]
\label{eq6}
\end{equation}
and where $N_{T}=1.23$, $R_{T}^{2}=6.5\mathrm{GeV}^{-2}$ \cite{lab19}.

By integrating over $\Delta$, (3) can be given as:
\begin{equation}
\begin{aligned}
&\frac{d\sigma_{\gamma A \rightarrow J/\psi A'}}{d^{2}b}=\int\limits_{0}^{1}dz\int\frac{d^{2}r}{4\pi}\Phi_{\gamma}{(\mathbf{r},z)}\\
&\times \int\limits_{0}^{1}dz'\int\frac{d^{2}r'}{4\pi}\Phi_{\gamma}^{\ast}{(\mathbf{r'},z')}[1-S^{\ast}(r')][1-S(r)]
\end{aligned}
\label{eq7}
\end{equation}

According to Mclerran-Venugopalan model\cite{lab11}, the $S$ factors are given by
\begin{equation}
S(r)\simeq\mathrm{exp}[-\frac{1}{8}Q_{s}^{2}r^{2}],
\label{eq8}
\end{equation}

\noindent where $Q_{s}$ is the gluon saturation scale function and its detailed form will be given in Eq.19.

(iii)In order to establish dipole-A interaction picture, DKLMT argued that the representation as shown in Fig.1.
The $J/\psi$ formed from c$\overline{c}$ is a color singlet. It is the particular dipole-nucleon inelastic collision
that converts the adjoint representation to a color singlet in the large-$N_{c}$ approximation.
The longitudinal coordinate $\xi$ (as shown in Fig.1 the distance from the front of the nucleus) indicates the point
where the particular inelastic interaction takes place. In order to keep the singlet intact, it is clear that
later interactions, occurring after the c$\overline{c}$ pair is in a singlet state, are purely elastic.

The interaction at $\xi$, responsible for the transition from a color octet state to a color singlet state,
can involve the anti-quark in both the amplitude and the conjugate amplitude.
Under the MV model evaluation employed here, the given interaction probability factor is:

\begin{equation}
\frac{Q_{s}^{2}\mathbf{r}\cdot\mathbf{r'}}{4T(b)}d\xi
\label{eq9}
\end{equation}

$T(b)$ depicts the nuclear thickness\cite{lab20} at given impact parameter $b$

\begin{equation}
T(b)=2\sqrt{R^{2}-b^{2}}\theta(R-b)
\label{eq10}
\end{equation}

\noindent where $R$ is the radius of target nucleus and $\theta$ is the step function.
The interactions occurring before $\xi$ can be taken into account.
With dipole separation $(\mathbf{r}-\mathbf{r'})/2$, each of these pieces can be treated as a dipole interaction
with the nucleus. This given the combined factor

\begin{equation}
\mathnormal{e}^{-\frac{1}{16}Q^{2}_{s}(\mathbf{r}-\mathbf{r'})^{2}(\xi/T(b))}
\label{eq11}
\end{equation}

\noindent where the $\xi/T(b)$ factor is for these interactions before the last inelastic interaction
at longitudinal coordinate $\xi$. As follows, we will provide these interactions
after the last inelastic interactions. The combined factor are given as follows:

\begin{equation}
\mathnormal{e}^{-\frac{1}{8}Q^{2}_{s}(\mathbf{r}^{2}+\mathbf{r'}^{2})(1-\xi/T(b))}
\label{eq12}
\end{equation}

The cross section of  $\mathit{g}A \rightarrow J/\psi X$ was given by \cite{lab12}:
\begin{equation}
\begin{aligned}
\frac{d\sigma_{ \mathit{g}A \rightarrow J/\psi X}}{d^{2}b}&=\int\limits_{0}^{1}dz\int\frac{d^{2}r}{4\pi}\Phi{(\mathbf{r},z)}
       \int\limits_{0}^{1}dz'\int\frac{d^{2}r'}{4\pi}\Phi^{\ast}{(\mathbf{r'},z')}\\
       &\times \int\limits_{0}^{T(b)}d\xi \frac{\mathbf{r\cdot r'}}{4T(b)}\mathrm{exp}\{-\frac{1}{16}Q^{2}_{s}(\mathbf{r}-\mathbf{r'})^{2}(\frac{\xi}{T(b)})\\
       &-\frac{1}{8}Q^{2}_{s}(\mathbf{r}^{2}+\mathbf{r'}^{2})(1-\frac{\xi}{T(b)})\}
\end{aligned}
\label{eq13}
\end{equation}

After this consideration, the cross-section distribution of $pA\rightarrow J/\psi X$ was provided \cite{lab12} as follows:
\begin{equation}
\begin{aligned}
\frac{d\sigma_{ \mathit{p}A \rightarrow J/\psi X}}{dyd^{2}b}&=x_{1}G(x_{1},m_{c}^{2})\int\limits_{0}^{1}dz\int\frac{d^{2}r}{4\pi}\Phi{(\mathbf{r},z)}\\
&\times\int\limits_{0}^{1}dz'\int\frac{d^{2}r'}{4\pi}\Phi^{\ast}{(\mathbf{r'},z')}\\
&\times\frac{4\mathbf{r\cdot r'}}{(\mathbf{r+r})^{2}}(e^{-\frac{1}{16}Q^{2}_{s}(\mathbf{r}-\mathbf{r'})^{2}}-e^{-\frac{1}{8}Q^{2}_{s}(\mathbf{r}^{2}+\mathbf{r'}^{2})})
\end{aligned}
\label{eq14}
\end{equation}
\noindent where

\begin{equation}
\begin{aligned}
\phi(\mathbf{r},z)&=\frac{g}{\pi\sqrt{2N_{c}}}\{ m_{c}^{2}K_{0}(m_{c}\mathbf{r})\\
&-[z^{2}+(1-z)^2]m_{c}K_{1}(m_{c}\mathbf{r})\partial_{r}\phi_{T}(\mathbf{r},z)\}
\end{aligned}
\label{eq15}
\end{equation}
\noindent that $K_{0}$ and $K_{1}$ are the modified Bessel functions.

According to DKLMT  and MV model(shown at Eq.9 and Eq.14), the productions of $J/\psi$ in $pA$ interactions
are independent of the longitudinal coordinate $\xi$, which means that at different location of $\xi$ the production probability is same.
We argued that the productions of $J/\psi$ in relativistic heavy-ion interactions should
rely on the longitudinal coordinate $\xi$ and impact parameter $b$.
When $b\ll R$ the dipole-nucleus collisions propagate through the whole nuclear thickness,
the probability for the inelastic scattering is thus proportional to the longitudinal coordinate.
This implies that the nuclear geometric effects play an important role to the $J/\psi$ production suppression.
At $b\approx R$, $J/\psi$ production in $dAu$ collisions becomes similar to
that in $pp$ collisions, the cold nuclear geometric effect is believed to be small and can be neglected.

In order to study the effects of dipole cold nuclear matter interaction at different point $\xi$ at different impact parameters inside the nucleus, we thus introduce a
nuclear geometric effect function $\mathit{f(\xi)}$ of the interaction at $\xi$ to account for the position dependence of the probability to form the $J/\psi$ under the
dipole-nucleus collisions as following:
\begin{equation}
\frac{Q_{s}^{2}\mathbf{r}\cdot\mathbf{r'}}{4T(b)}f(\xi)d\xi
\label{eq16}
\end{equation}

After introducing the nuclear geometric effect function $\mathit{f(\xi)}$ to take into account the different probability of inelastic interaction at different location $\xi$,
the cross section for $J/\psi$ production in $pA$ collisions becomes a function of the nuclear thickness function. The nuclear geometric effect function $\mathit{f(\xi)}$ is not a flat distribution of $\xi$ in our assumption,  the smaller the magnitude of $\xi$,  the larger the value of $\mathit{f(\xi)}$. The form of the nuclear matter coherent function
is assumed to be a likely Gaussian form:
\begin{equation}
f(\xi)=\alpha e^{-\beta(\frac{\xi}{T(b)})^{2}}
\label{eq17}
\end{equation}

\noindent where $\alpha$ is a normalization factor, $\beta$ is an adjustable parameter which can be determined from experimental data.

In order to calculate the nuclear modification factor $R_{dA}$ at different centrality and at different rapidity regions, it is necessary to describe the nuclear geometry feature properly.
The relation between impact parameter $b$ and the number of participant ($N_{part}$) in $dAu$ collisions is already derived in Ref.\cite{lab21} and given by
\begin{equation}
b=R_{Au}\sqrt{1-\frac{(N_{part}-2)^{2}}{A^{2}[1-(1-\frac{R^{2}_{d}}{R_{Au}^{2}})^{3/2}]^{2}}}
\label{eq18}
\end{equation}

\noindent where $A$ is the number of target nucleon of gold, $R_{Au}$ the radius of target gold nucleus and $R_{d}$ the radius of the projectile deuterium nucleus.

The DHJ model \cite{lab17} has improved the KKT model \cite{lab22,lab23} by taking into account the change in the anomalous dimension of the gluon distribution
function due to the presence of the saturation boundary \cite{lab24} and also some higher order effects.
The DHJ model performs the numerical calculations of the dipole scattering amplitude \cite{lab17} as follows:

\begin{equation}
N_{A}(\mathbf{r},y)=1-\mathrm{exp}\{-\frac{1}{4}(r^{2}Q^{2}_{s})^{\gamma}\}
\label{eq19}
\end{equation}

The gluon saturation scale is given by
\begin{equation}
Q_{s}^{2}=\Lambda^{2}A^{1/3}e^{\lambda y}=0.13 \mathrm{GeV}^{2}e^{\lambda y}N_{coll},
\label{eq20}
\end{equation}

\noindent and the parameters $\gamma$ is the anomalous dimension
\begin{equation}
\gamma=\gamma_{s}+(1-\gamma_{s})\frac{\ln(m^{2}/Q_{s}^{2})}{\lambda Y+\ln(m^{2}/Q_{s}^{2})+d\sqrt{Y}}
\label{eq21}
\end{equation}
 where $\gamma_{s}=0.628$ is implied by theoretical arguments\cite{lab25} and d=1.2, $Y=\ln(1/x)$, $x=me^{-y}\sqrt{s}$, $\Lambda=0.6 \mathrm{GeV}$, $\lambda=0.3$ are fixed by DIS data \cite{lab26,lab27}. Besides DHJ model, some model \cite{lab28} also used
the anomalous dimension of the gluon distribution function to study dipole scattering amplitude.

After the consideration of geometric modification of DKLMT model, we provide the cross section as follow:
\begin{equation}
\begin{aligned}
&\frac{d\sigma_{ \mathit{g}A \rightarrow J/\psi X}}{d^{2}b}=\int\limits_{0}^{1}dz\int\frac{d^{2}r}{4\pi}\Phi{(\mathbf{r},z)}
\int\limits_{0}^{1}dz'\int\frac{d^{2}r'}{4\pi}\Phi^{\ast}{(\mathbf{r'},z')}\\
       &\times \int\limits_{0}^{T(b)}d\xi \frac{\mathbf{r\cdot r'}}{4T(b)}\mathrm{exp}\{-\beta(\frac{\xi}{(T(b))})^{2}-\frac{1}{16}Q^{2}_{s}(\mathbf{r}-\mathbf{r'})^{2}(\frac{\xi}{T(b)})\\
       &-\frac{1}{8}Q^{2}_{s}(\mathbf{r}^{2}+\mathbf{r'}^{2})(1-\frac{\xi}{T(b)})\}
\end{aligned}
\label{eq22}
\end{equation}

The cross section of  $pA\rightarrow J/\psi X$ is given by
\begin{equation}
\begin{aligned}
&\frac{d\sigma_{ \mathit{p}A \rightarrow J/\psi X}}{dyd^{2}b}=x_{1}G(x_{1},m_{c}^{2})\frac{\alpha}{2}\sqrt{\frac{\pi}{\beta}}\int_{0}^{1}dz
\int\frac{d^{2}r}{4\pi}\Phi{(\mathbf{r},z)}\\
&\times\int\limits_{0}^{1}dz'\int\frac{d^{2}r'}{4\pi}\Phi^{\ast}{(\mathbf{r'},z')}\frac{Q^{2}_{s}\mathbf{r\cdot r'}}{4}
e^{-\frac{1}{8}Q^{2}_{s}(\mathbf{r}+\mathbf{r'})^{2}+\frac{1}{32^{2}\beta}Q^{4}_{s}(\mathbf{r}+\mathbf{r'})^{4}} \\
&\times\lbrace(\Phi(\frac{Q^{2}_{s}(\mathbf{r}+\mathbf{r'})^{2}}{32\sqrt{\beta}})+\Phi(\sqrt{\beta}
(1-\frac{Q^{2}_{s}(\mathbf{r}+\mathbf{r'})^{2}}{32\beta}))\rbrace\\
\end{aligned}
\label{eq23}
\end{equation}

\noindent where $\Phi(\mu)=\int\limits_{0}^{\mu}e^{-x^{2}}dx$ is the error function. Compared with
Eq.14 given by DKLMT model, Eq.23 has a different form by introducing the Gaussian function $f(\xi)$ of nuclear geometrical effect.
One can found that the cross section of  $pA\rightarrow J/\psi X$ is sensitive to the Gaussian form and the parameter $\beta$. By fit
the experimental results of RHIC, the given $\beta$ is  9.

After consideration of DHJ model, the cross section is given as follow:
\begin{equation}
\begin{aligned}
&\frac{d\sigma_{ \mathit{p}A \rightarrow J/\psi X}}{dyd^{2}b}=x_{1}G(x_{1},m_{c}^{2})\frac{\alpha}{2}\sqrt{\frac{\pi}{\beta}}\int_{0}^{1}dz
\int\frac{d^{2}r}{4\pi}\Phi{(\mathbf{r},z)}\\
&\times\int\limits_{0}^{1}dz'\int\frac{d^{2}r'}{4\pi}\Phi^{\ast}{(\mathbf{r'},z')}\frac{(Q_{s}\mathbf{r})^{\gamma}\cdot (Q_{s}\mathbf{r'})^{\gamma}}{4}\\
&\times exp\{{-\frac{1}{8}(Q_{s}(\mathbf{r}+\mathbf{r'}))^{2\gamma}+\frac{1}{32^{2}\beta}(Q_{s}\cdot(\mathbf{r}+\mathbf{r'}))^{4\gamma}}\} \\
&\times\lbrace(\Phi(\frac{(Q_{s}(\mathbf{r}+\mathbf{r'}))^{2\gamma}}{32\sqrt{\beta}})+\Phi(\sqrt{\beta}
(1-\frac{(Q_{s}(\mathbf{r}+\mathbf{r'}))^{2\gamma}}{32\beta}))\rbrace\\
\end{aligned}
\label{eq24}
\end{equation}

\section{Calculations and results}
In this section, we calculate nuclear modification factor and compare the results to the experimental measurements in 200 GeV
$dAu$ collisions at RHIC \cite{lab13,lab14,lab15,lab16}. To study the nuclear matter effect in $dAu$ collisions, we
recall the definition of nuclear modification factor (NMF)
\begin{equation}
R_{dA}=\frac{d\sigma_{J/\psi}^{dAu}/dy}{<N_{coll}>d\sigma_{J/\psi}^{pp}/dy}
\label{eq25}
\end{equation}

\noindent where $N_{coll}$ is the number of binary nucleon-nucleon collisions.

The NMF results of our M-DKLMT model in $dAu$ collisions at RHIC are shown in Figs. 2, 3 and 4.
\begin{figure*}[!htbp]
\centering \resizebox{0.555\textwidth}{!}{
\includegraphics{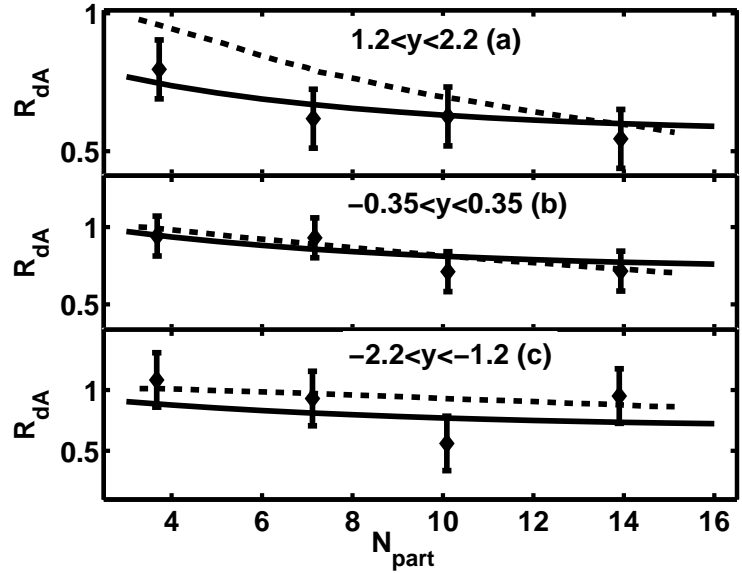}}
\caption{The dependencies of nuclear modification factors $R_{dA}$ on the number of participants at different rapidity regions: (a) for $1.2 < y < 2.2$,
(b) for $-0.35 < y < 0.35$ and (c) for $-2.2 < y < -1.2$. The experimental results come
from \cite{lab13,lab14,lab15,lab16}. The solid lines are our calculation results and the dashed
lines are the results from DKLMT model.}
\label{fig2} 
\end{figure*}

Figure 2 shows the dependencies of nuclear modification factors $R_{dA}$ on the number of participants at different rapidity regions.
One can find that when introducing the Gaussian geometric effect function $f(\xi)$, a strong suppression of $R_{dA}$
at small $N_{part}$ is shown in our M-DKLMT model. The results shown in Fig.2 indicate that our M-DKLMT model describes the experimental data better than that of the DKLMT model,
especially at the forward rapidity of $1.2 < y < 2.2$.

\begin{figure*}[!htbp]
\centering \resizebox{0.65\textwidth}{!}{
\includegraphics{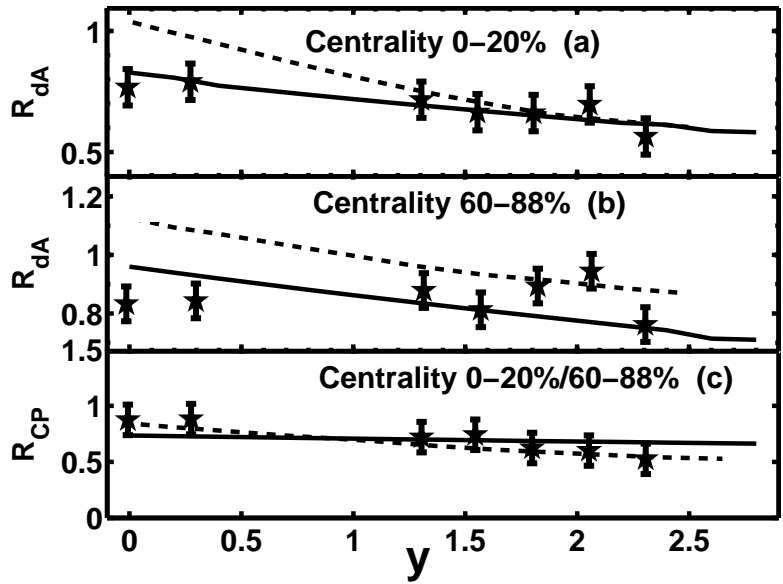}}
\caption{The rapidity dependencies of $R_{dAu}$ on different collision centralities: (a) $0-20\%$,
(b) $60-88\%$ and (c) for $R_{CP}$ the ratio of $0-20\%/60-88\%$. The solid lines are our calculation
results and the dashed lines are the results from the CGC model \cite{lab29}.}
\label{fig3} 
\end{figure*}

The rapidity dependencies of $R_{dAu}$ for different collision centralities are shown in Fig.3. Comparing with the results of DKLMT,
we find that our M-DKLMT model results in a stronger suppression at mid-rapidity region.
In addition, our M-DKLMT model describes the experimental data better than that of the DKLMT model, especially for the central collisions as shown in Fig.3(a).
When considering the nuclear geometric effect, we find that  the nuclear thickness T(b) of central collisions is large, the nuclear medium effect is obvious,
which can reflect one of the characteristics of cold nuclear geometric effects.

\begin{figure*}[!htbp]
\centering \resizebox{0.65\textwidth}{!}{
\includegraphics{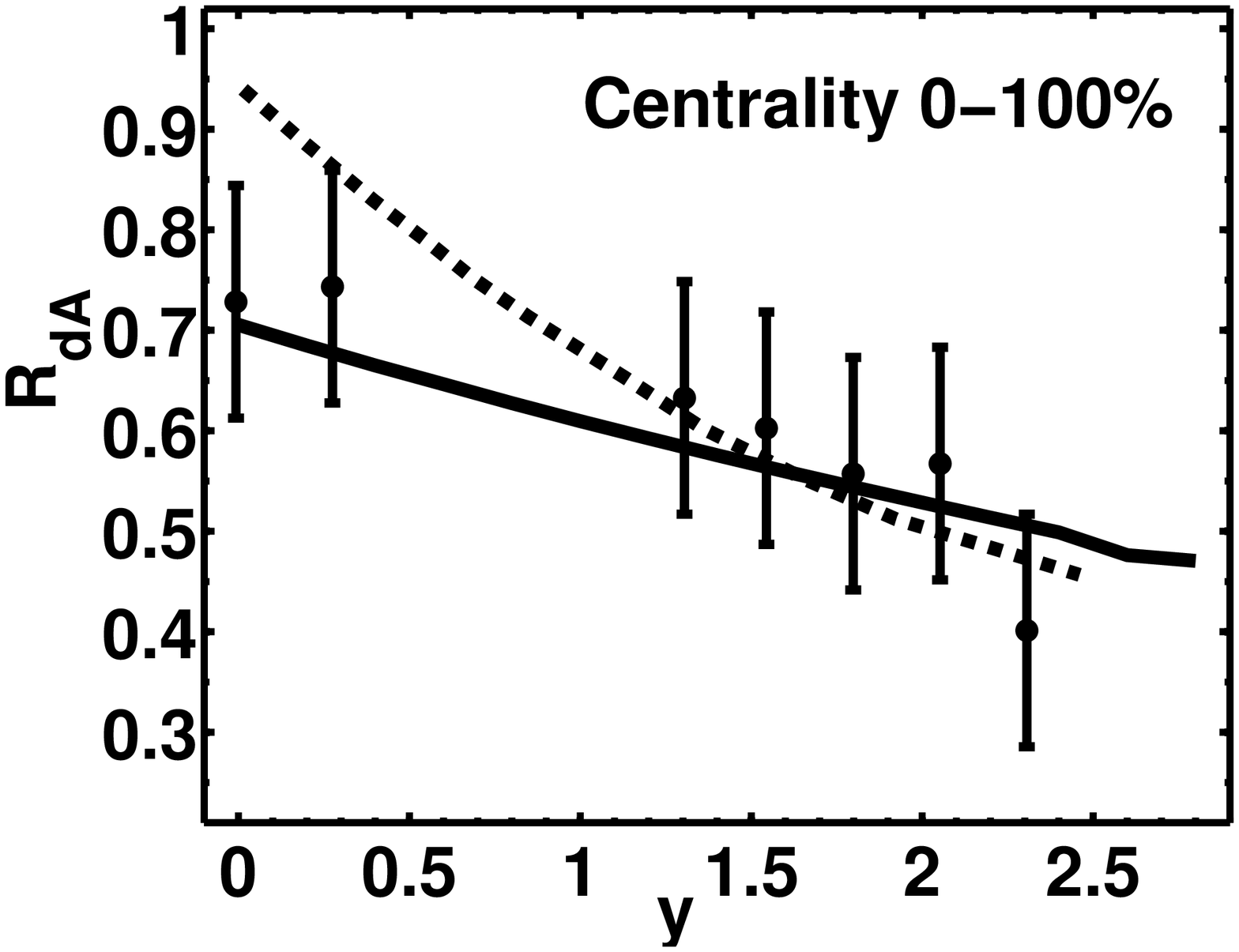}}
\caption{The rapidity dependencies of nuclear modification factors for centrality of $0-100\%$.
The solid line indicates our calculation result and the dashed line shows the result
from the CGC model \cite{lab33}.}
\label{fig4} 
\end{figure*}

Figure 4 shows the rapidity dependencies of the nuclear modification factors for centrality of $0-100\%$. The solid line shows our calculation
result and the dashed line is the result from the DKLMT model calculation. Comparing with the DKLMT model, the M-DKLMT model calculation shows
good agreement with experimental data.

\section{Conclusions and Discussions}
The studies of $p(d)A$ collisions at different energies were motivated in order to understand cold nuclear matter effects \cite{lab30, lab31, lab32, lab33}. These CNM effects can modify $J/\psi$ production in $pA$ collisions as compared to pp collisions where in both cases a QGP is believed to be absent. CNM effects that were often considered include nuclear modification of the parton distributions in nuclei (nPDFs), break up of the $J/\psi$ precursor  state in the cold nucleus, parton transverse momentum broadening in traversing the cold nucleus, and initial state parton energy loss \cite{lab30, lab31}.  It has been hoped that CNM effects and hot matter effects can be factorized, so that CNM effects can be measured in $p(d)A$ collisions and accounted for when analyzing heavy ion collisions data to extract hot dense medium effects.

Ref.34 make a research of the production of heavy quarkonium states in high energy proton-nucleus collisions and systematically included both small $x$
evolution and multiple scattering effects on heavy quark pair production within the Color Glass Condensate (CGC) framework. It was observed \cite{lab34}
that the production of color singlet heavy quark pairs is sensitive to both ¡°quad-rupole¡± and ¡°dipole¡± Wilson line correlators, whose energy evolution is
described by the Balitsky-JIMWLK equations. In contrast, the color octet channel is sensitive to dipole correlators alone.
In a quasi-classical approximation, their results for the color singlet channel reduce to those of Dominguez et al. \cite{lab12}.

In this paper we developed a modified DKLMT model to describe the cold nuclear matter effects on the color singlet $J/\psi$ productions in $dAu$ collisions at RHIC. In order to describe the centrality and rapidity dependencies of nuclear modification factor ($R_{dA}$) for $J/\psi$ productions at RHIC, the nuclear geometric effect function $f(\xi)$ and the relationship between impact parameter $b$ and the number of participants ($N_{part}$) in $dAu$ collisions are introduced to the M-DKLMT model. The nuclear geometric effect function $f(\xi)$ is mainly to account for the different interaction probability at different location $\xi$. It is realized that the nuclear geometric effect function $f(\xi)$ is not uniform but has a larger value at smaller $\xi$.

One can find that the M-DKLMT model introduce a stronger suppression at small $N_{part}$ and at mid-rapidity region by comparing with DKLMT model. The M-DKLMT model can describe the experimental data better than that of the DKLMT model, especially for the forward rapidity region $1.2 < y < 2.2$ and for the central collisions.

\section{Acknowledgments}
This work was supported by the National Natural Science Foundation of China (11475068, 11247021, 11375071, 10975091), the National Basic
Research Program of China (2013CB837803) , the Basic Research Program of CCNU (CCNU13F026),  and Key Laboratory foundation of
Quark and Lepton Physics (Central China Normal University)(QLPL2014P01).

{}
\end{document}